\documentclass[
aps,
prl,
reprint,
groupedaddress
]{revtex4-2}

\usepackage{amsmath}
\usepackage{amsfonts}
\usepackage{graphicx}
\usepackage{amssymb}
\usepackage{textcomp}
\usepackage{bm}
\usepackage{color}
\usepackage{dsfont}

\usepackage{pgfplots}

\usepackage{siunitx}

\usepackage{mathtools}

\usetikzlibrary{arrows,shapes,backgrounds,calc,
    positioning,
    intersections}

\definecolor{Blue}{cmyk}{1,0.6,0,0.05}
\definecolor{Red}{cmyk}{0.04,0.87,0.89,0}
\definecolor{Green}{cmyk}{1,0.,1,0}
\definecolor{Yellow}{cmyk}{0,0.2,1,0}
\definecolor{Orange}{cmyk}{0.0,0.69,1,0}
\definecolor{Orange2}{cmyk}{0.0,0.6,1,0.00}
\definecolor{SoftCyan}{cmyk}{0.48,0.,0.07,0.11}

\definecolor{BlenderOrange}{rgb}{1, 0.5, 0.168}
\definecolor{BlenderOrangeDark}{rgb}{1., 0.43, 0.27} 

\definecolor{LaserBlue}{rgb}{0.4,0,1}
\definecolor{LaserMOT}{rgb}{1,0.372,0}
\definecolor{LaserJseven}{rgb}{1,0,0}
\definecolor{LaserJeight}{rgb}{0.35,0,0}

\pgfplotsset{colormap={CM}{color=(white) color=(Blue!50!white) color=(Blue)  color=(Blue!75!black) color=(Blue!50!black)  color=(Blue!25!black) color=(black)}}

\pgfplotsset{colormap={RedToBlue}{color=(Red)  color=(white!50!black) color=(Blue)}}

\pgfplotsset{colormap={CM2}{color=(white) color=(Blue!33!white) color=(Blue!67!white) color=(Blue)  color=(Blue!75!black) color=(Blue!50!black)  color=(Blue!25!black) color=(black)}}
\pgfplotsset{colormap={CM3}{color=(white)  color=(Blue)  color=(black)}}
\pgfplotsset{colormap={CM4}{color=(white)  color=(Red)  color=(black)}}

\DeclareMathOperator{\acos}{acos}

\newcommand{\fig}[1]{Fig.\,\ref{#1}}

\newcommand{\eqP}[1]{(\ref{#1})}

\newcommand{\bitem}{\begin{itemize}}
\newcommand{\eitem}{\end{itemize}}

\newcommand{\bti}{\begin{tikzpicture}}
\newcommand{\eti}{\end{tikzpicture}}

\newcommand{\ket}[1]{\left| #1 \right>} 
\newcommand{\bra}[1]{\left< #1 \right|} 

\newcommand{\Er}{E_{\text{r}}}

\newcommand{\dd}{\text{d}}

\newcommand{\hc}{\text{hc}}
\newcommand{\I}{\text{i}}

\newcommand{\E}{\text{e}}

\newcommand{\ebold}{\mathbf{e}}

\newcommand{\Bbold}{\mathbf{B}}

\newcommand{\Veff}{V_{\text{eff}}}

\newcommand{\bas}{\begin{align}}
\newcommand{\eas}{\end{align}}

\newcommand{\bc}{\begin{center}}
\newcommand{\ec}{\end{center}}

\pgfplotsset{every axis/.append style={
  x tick label style={font=\small,yshift=0.5mm},
  y tick label style={font=\small,xshift=0.5mm},
  label style={font=\small},
  xlabel style={yshift=2.5mm},
  ylabel style={yshift=-7mm},
}}

\newcommand{\vrec}{v_{\text{rec}}}

\newcommand{\Mod}[1]{\ (\mathrm{mod}\ #1)}

\newcommand{\X}{x}
\newcommand{\vx}{v_\X}
\newcommand{\Fx}{F_\X}

\definecolor{col0}{HTML}{520BDE}
\definecolor{col1}{HTML}{F51400}
\definecolor{col2}{HTML}{E0A90B}
\definecolor{col3}{HTML}{26FF00}

\colorlet{colLaser1}{Red}
\colorlet{colLaser2b}{Red!70!black}
\colorlet{colLaser2a}{Red!70!white}

\colorlet{col0}{Blue}
\colorlet{col1}{Red}
\colorlet{col2}{Green}

\newlength{\OneColumnPRLWidth}
\newcommand{\phia}{\varphi_a}
\newcommand{\phib}{\varphi_b}

\DeclareSIUnit\gauss{G}

\newcommand{\kL}{k}
\newcommand{\lmag}{\ell_{\text{mag}}}

\renewcommand{\r}{y}
\newcommand{\Lr}{Y}
\newcommand{\deltaZ}{\delta_{\text{Z}}}

\begin{document}

 \title{
Laughlin's topological charge pump in an atomic  Hall cylinder
 }

 \author{Aurélien Fabre}
 \author{Jean-Baptiste Bouhiron}
 \author{Tanish Satoor}
\author{Raphael Lopes}
\author{Sylvain Nascimbene}
\email{sylvain.nascimbene@lkb.ens.fr}
 \affiliation{Laboratoire Kastler Brossel,  Coll\`ege de France, CNRS, ENS-PSL University, Sorbonne Universit\'e, 11 Place Marcelin Berthelot, 75005 Paris, France}
 \date{\today}

  \begin{abstract}
The quantum Hall effect occuring in two-dimensional electron gases was first explained by Laughlin, who envisioned a thought experiment that laid the groundwork for our understanding of topological quantum matter. His proposal is based on a quantum Hall cylinder periodically driven by an axial magnetic field, resulting in the quantized motion of electrons. We realize this milestone experiment with an ultracold gas of dysprosium atoms, the cyclic dimension being encoded in the electronic spin and the axial field controlled by the phases of laser-induced spin-orbit couplings. Our experiment provides a straightforward manifestation of the non-trivial topology of quantum Hall insulators, and could be generalized to strongly-correlated topological systems.
 \end{abstract}
 
 \maketitle

\begin{figure}[!t]
 \begin{center}
 \includegraphics[
  trim={2mm 5mm 0 0.cm},width=6cm
 ]{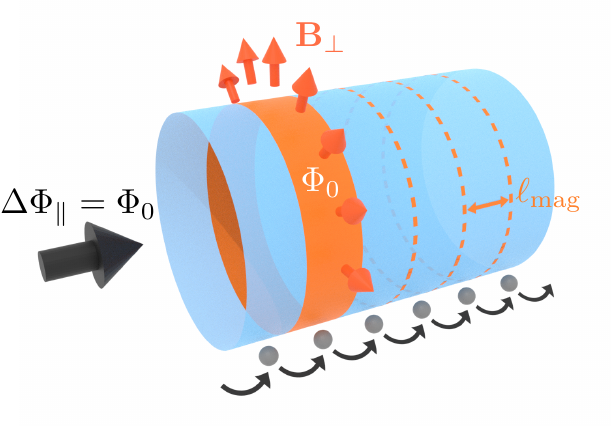}
 \end{center}
 \caption{
\textbf{Laughlin's thought experiment.} Scheme of a two-dimensional electronic system in a cylindrical geometry, with a radial magnetic field $\Bbold_\perp$ producing a quantum Hall effect. The orange area, pierced by one magnetic flux quantum $\Phi_0$, defines the length $\lmag$ of the magnetic unit cell -- each cell being filled with one electron in a quantum Hall insulator. Laughlin's thought experiment consists in performing an adiabatic cycle by threading one flux quantum $\Delta\Phi_\parallel=\Phi_0$ through the cylinder. The cycle shifts electron occupations by one unit cell, such that a single electron is pumped from one edge to the other, or equivalently the  center-of-mass position is displaced by $\lmag$. 
 \label{fig_scheme}}
 \end{figure}

The quantization of  Hall conductance observed in two-dimensional electronic systems subjected to a perpendicular magnetic field \cite{klitzing_new_1980} is intimately linked to the non-trivial topology of Bloch bands \cite{thouless_quantized_1982} and the occurence of chiral edge modes protected from backscattering \cite{halperin_quantized_1982}. The first step in its undertanding was provided by  Laughlin, who gave an elegant argument by considering  a Hall system in a cylindrical geometry (\fig{fig_scheme}) \cite{laughlin_quantized_1981}. Besides the radial magnetic field $\Bbold_\perp$ yielding the Hall effect, this geometry authorizes an axial field $\Bbold_\parallel$, which does not pierce the surface but threads the cylinder with a flux $\Phi_\parallel$. Varying the flux $\Phi_\parallel$ controls a quantized  electronic motion along the tube, which is directly linked to the underlying band topology.  Such quantization of transport was later generalized by Thouless to any physical system subjected to a slow periodic deformation  \cite{thouless_quantization_1983}, as implemented in electronic quantum dots \cite{switkes_adiabatic_1999,watson_experimental_2003}, photonic waveguides \cite{kraus_topological_2012} and ultracold atomic gases \cite{lohse_thouless_2016,nakajima_topological_2016}.

So far, the topology of magnetic Bloch bands has been revealed in planar systems only,   by measuring the quantization of transverse response \cite{klitzing_new_1980,dean_hofstadters_2013,ponomarenko_cloning_2013,aidelsburger_measuring_2015} or observing chiral ballistic edge modes \cite{wang_observation_2009,hafezi_imaging_2013,rechtsman_photonic_2013}. The realization of Laughlin's pump experiment requires engineering periodic boundary conditions, which is challenging when using genuine spatial dimensions. The concept of a synthetic dimension encoded in an internal degree of freedom provides an interesting alternative  \cite{celi_synthetic_2014,mancini_observation_2015,stuhl_visualizing_2015}, which recently led to the realization of synthetic Hall cylinders \cite{han_band_2019,li2018bose,liang_coherence_2021}. 

In this work, we use an ultracold gas of $^{162}$Dy atoms to engineer a  Hall cylinder whose azimuthal coordinate is  encoded in the electronic spin $J=8$ \cite{chalopin_probing_2020}. We manipulate the spin using coherent optical transitions,  such that a triplet of internal states coupled in a cyclic manner emerges at low energy, leading to an effective cylindrical geometry \cite{fabre_2021}. The exchange of momentum between light and atoms leads to a spin-orbit coupling that mimics a radial magnetic field $\Bbold_\perp$ \cite{dalibard_colloquium_2011}. The phases of the laser electric fields also control an effective axial  field $\Bbold_\parallel$, which we use to implement Laughlin's thought experiment and reveal the underlying topology. The topological character of the ground Bloch band manifests as well in a complementary pump experiment driven by Bloch oscillations.

\begin{figure}[!t]
\begin{center}
\includegraphics[
 trim={2mm 4mm 0 0.cm},width=9.4cm
]{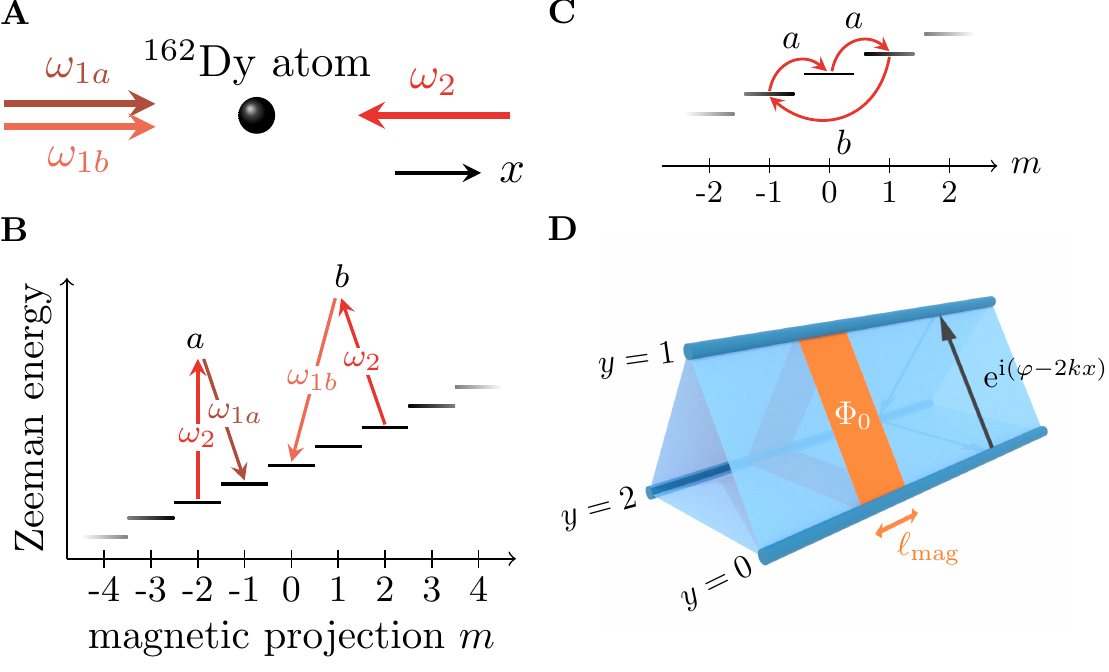}
\end{center}
\caption{
\textbf{Emerging quantum Hall cylinder.}
(\textbf{A}) Sketch of the laser configuration involving two beams counter-propagating along $\X$ and sent on a thermal sample of dysprosium atoms -- one beam having two frequency components. (\textbf{B}) Scheme of the two-photon optical transitions resonantly driving first- and second-order spin transitions, labelled $a$ and $b$, respectively. (\textbf{C}) Representation of a non-trivial 3-cycle between magnetic sub-levels induced by the light couplings.  (\textbf{D}) Scheme of the Hall cylinder dynamics emerging at low energy, involving  three spin states $\ket{\r}$ (with $\r=0,1,2$). 
The hopping amplitudes have a complex phase $\varphi-2kx$, where $2\hbar k$ plays the role of a radial magnetic field 
$B_\perp$ and $\varphi$ is linked to an axial field $B_\parallel$. 
The orange area, of length $\lmag=\lambda/6$ is threaded by one unit of magnetic flux quantum $\Phi_0$.
\label{fig_model}}
\end{figure}

In our experimental protocol, we apply a magnetic field in order to lift the degeneracy between the magnetic sub-levels $m$ (with $-J\leq m\leq J$ and integer $m$). Spin transitions of first and second order, \emph{i.e.} \mbox{$\Delta m=\pm1$ and $\pm2$}, are induced by resonant two-photon optical transitions, using a pair of laser beams counter-propagating along $\X$ (\fig{fig_model}A)   \cite{Note1}. The configuration of laser frequencies is chosen such that the atoms undergo a momentum kick $-2\hbar k$ upon either resonant process $m\rightarrow m+1$ or $m\rightarrow m-2$ shown in \fig{fig_model}B.  Here, $k=2\pi/\lambda$ is the photon momentum for the laser wavelength $\lambda=\SI{626.1}{\nano\meter}$. The resulting spin-orbit coupling breaks continuous translation symmetry, but conserves the quasi-momentum $q=M\vx/\hbar+2k m\Mod{6k}$, defined over the magnetic Brillouin zone  $-3k\leq q<3k$, where $M$ and $v_x$ are the atomic mass and velocity.
The atom dynamics is  described by the Hamiltonian
\begin{align}
H&=\frac{1}{2}Mv_x^2+V,\label{eq_H}\\
V&=-\mathcal{T}_\r\,\E^{-2\I kx}+\hc,\quad \mathcal{T}_\r=t_a\E^{\I \phia} \frac{J_+}{J}+t_b\E^{\I \phib}\frac{J_-^2}{J^2},\label{eq_V}
\end{align}
where $J_+$ and $J_-$ are the spin ladder operators, and $t_a,t_b>0$ are the strengths of the first and second-order transitions.
The phase difference $\phia-\phib$ can be gauged away using a suitable spin rotation, such that we  retain hereafter a single  phase $\varphi\equiv\phia=\phib$.

The combination of the two types of transitions induces non-trivial 3-cycles $m\rightarrow m+1\rightarrow m+2\rightarrow m$ (\fig{fig_model}B), with chiral dynamics in the cyclic variable $\r=m\Mod{3}$ -- each step increasing $\r$ by one unit. As explained in \cite{fabre_2021}, one expects the emergence of a closed subsystem at low energy, spanned by three spin states $\ket{\r}$, with $\r=0,1,2$ and where $\ket{ \r}$  expands on projection states $\ket{m}$ with $m=\r\Mod{3}$ only. The $\ket{\r}$ states will be interpreted in the following as position eigenstates along a cyclic synthetic dimension of length $\Lr=3$. The operator $\mathcal{T}_y$ involved in the spin coupling (\ref{eq_V}) then acts as a translation $\mathcal{T}_y\ket{y}=t\ket{y+1}$, with a hopping amplitude $t=  t_a+ t_b$. The low-energy spin dynamics  is described by the effective potential
\begin{equation}
\Veff= -t\sum_{\r=0}^2  \left(\E^{\I(\varphi-2k\X)}\ket{\r+1}\bra{\r}+\hc\right).\label{eq_Veff}
\end{equation}
Together with the kinetic energy $\tfrac{1}{2}M v_x^2$, it describes the motion of a particle on a cylinder discretized along its circumference (see \fig{fig_model}C). The complex phase $2k\X$ mimics the Aharonov-Bohm phase associated with a radial magnetic field $B_\perp=2\hbar k$  (assuming a particle charge $q=-1$). It defines a magnetic length $\lmag=\lambda/6$, such that the magnetic flux $\Phi_\perp=\lmag \Lr B_\perp$ through  a portion of cylinder of length $\lmag$ equals the flux quantum $\Phi_0=h/|q|$.

\begin{figure*}[!t]
\begin{center}
 \includegraphics[
 draft=false,scale=1,
 trim={3mm 5mm 0 0.cm},
]{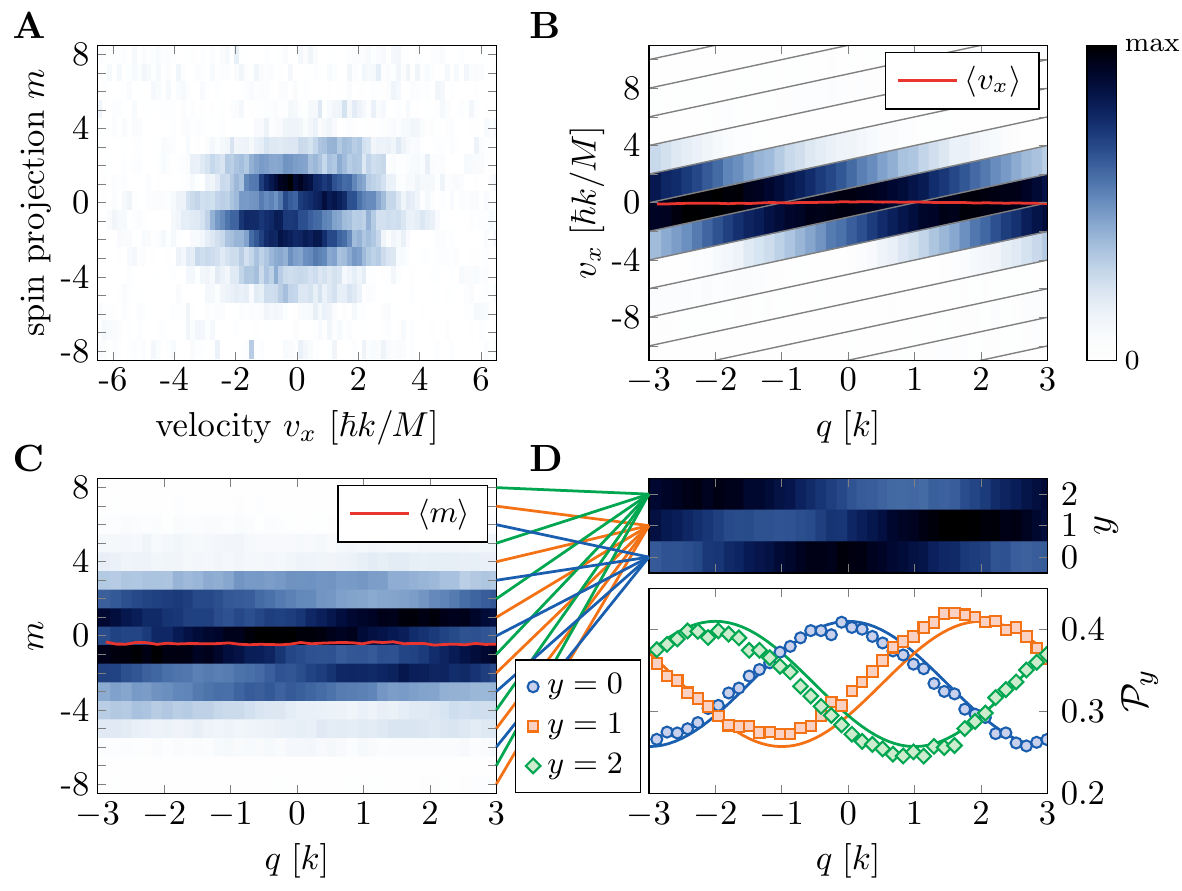}
\end{center}
\caption{
\textbf{Ground band characterization.} (\textbf{A}) Spin-resolved velocity distribution measured for a gas of  mean quasi-momentum $\langle q\rangle\simeq2k$. 
(\textbf{B}) Distribution of discrete velocity components $v_x=\hbar(q+2k p)/M$ (with integer $p$) for states of quasi-momentum $q$. The red line shows the mean velocity $\langle \vx\rangle$. (\textbf{C}) Spin projection probabilities $\Pi_m$ measured as a function of $q$. The red line stands for the mean spin projection $\langle m\rangle$. (\textbf{D}) Probabilities $\mathcal{P}_\r$ of projection on $\r=m\Mod{3}$. The blue circles, orange squares and green diamonds correspond to $y=0$, 1 and 2, respectively. Statistical error bars, computed from a bootstrap random sampling analysis, are smaller than the symbols. The lines are calculated from the expected band structure. 
\label{fig_ground_band}}
\end{figure*}

Experimentally, we use a gas of about $ 4 \times 10^4$ atoms, initially prepared at a temperature $T=\SI{0.54(3)}{\micro\kelvin}$, such that the thermal momentum width $\sigma_q\simeq k$ is much smaller than the Brillouin zone extent, and interaction effects can be neglected on the timescale of our experiments. The atoms are adiabatically loaded in the ground Bloch band by ramping the light coupling parameters, and the mean quasi-momentum $\langle q\rangle$ is controlled by applying a weak force $F_x$ after the loading  (see the Supplementary Materials \cite{Note1}). We simultaneously probe the distribution of velocity $v_x$ and spin projection $m$. For this, we measure the  atom distribution after time-of-flight in the presence of a magnetic field gradient, which separates the different magnetic sub-levels. A typical spin-resolved velocity distribution is shown in \fig{fig_ground_band}A.

The velocity distribution, plotted in \fig{fig_ground_band}B as a function of $q$, exhibits a period $2k$, similar to the case of a simple $\lambda/2$-lattice.  The mean velocity $\langle \vx\rangle$,  shown as a red line, remains close to zero. Since it is linked to the slope of the ground-band energy $\partial_q E_0(q)=\hbar\langle \vx\rangle$, this shows that the band is quasi-flat.
In fact, the band's flatness in protected from pertubations, such as external magnetic field fluctuations, by the zero net magnetization of the $\ket{y}$ spin states --  a similar effect has been used in another implementation of a Hall cylinder using dynamical decoupling techniques \cite{liang_coherence_2021}. 

The probabilities $\Pi_m$ of projection on each sub-level $m$ reveal a longer periodicity $6k$ (\fig{fig_ground_band}C), corresponding to the full extent of the magnetic Brillouin zone. It experimentally confirms  the spatial separation of magnetic orbitals $\lmag=2\pi/(6k)=\lambda/6$ introduced above. The $\Pi_m$ measurements also give access to the probabilities $\mathcal{P}_\r$ of projection on the synthetic coordinate $\r$, by summing the $\Pi_m$'s with $m=\r\Mod{3}$ (\fig{fig_ground_band}D). The $q$-variation of these distributions reveals a chirality typical of the Hall effect: when  increasing the momentum by $2k$, the $\mathcal{P}_\r$ distributions cycle along the synthetic dimension in a directional manner, as  $\mathcal{P}_\r\rightarrow \mathcal{P}_{\r+1}$
 \cite{yan_emergent_2018,anderson_realization_2020}. We stress that such a drift does not occur on the mean spin projection $\langle m \rangle$, which remains close to zero (red line in \fig{fig_ground_band}C).

\begin{figure*}[!t!]
\begin{center}
 \includegraphics[
 draft=false,scale=1,
 trim={5mm 5mm 0 0.cm},
]{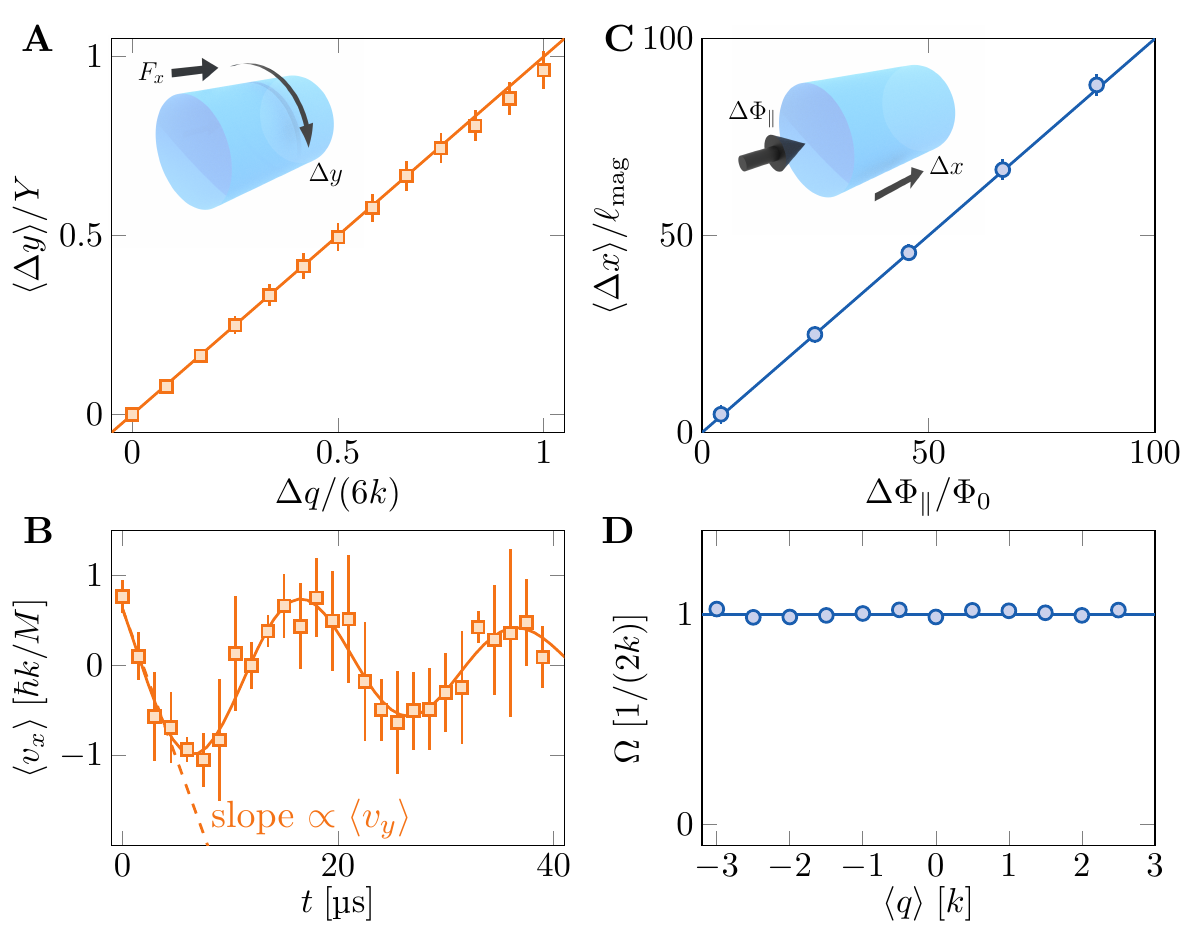}
\end{center}
\caption{
\textbf{Topological charge pumps.} (\textbf{A}) Center-of-mass displacement $\langle\Delta\r\rangle$ as a function of the quasi-momentum shift $\Delta q$ induced by a force $\Fx$ (orange squares), together with a linear fit (orange line). (\textbf{B}) Evolution of the mean velocity $\langle \vx\rangle$ immediately after switching off the force $\Fx$ (orange squares), fitted with a damped sine (solid line). The velocity $\langle v_\r\rangle$ is obtained from the initial slope of the fit (dashed line). (\textbf{C}) Displacement of the center of mass $\langle \Delta\X\rangle$ induced by an axial magnetic flux $\Phi_\parallel$ and averaged over the magnetic Brillouin zone (blue cicles). The blue line is a linear fit.  (\textbf{D}) Berry curvature $\Omega$ measured as a function of the mean quasi-momentum $\langle q\rangle$ (blue circles). The solid line is the expected Berry curvature, which is not distinguishable from the constant value $\Omega(q)=1/(2k)$.   
\label{fig_pump}}
\end{figure*}

The adiabatic $y$-drift occuring during Bloch oscillations provides a first insight into the topological character of the lowest energy band -- similar to the quantized flow of Wannier function charge centers  in  Chern insulators \cite{taherinejad_wannier_2014}. To quantify this drift, we cannot rely on the mean $y$ position, which is ill-defined for a cyclic dimension \cite{lynch_quantum_1995}. Instead, it is reconstructed by integrating the anomalous velocity $\langle v_\r\rangle\equiv\partial_\varphi H/\hbar$ induced by the force $\Fx$ driving the Bloch oscillation. For this purpose, we conduct a separate experiment, in which we suddenly switch off the force $\Fx$, such that the center-of-mass undergoes a cyclotron oscillation, with the $x$- and $\r$-velocities oscillating in quadrature. More precisely, the rate of change of the $x$-velocity gives access to the $y$-velocity, via the exact relation
 \[
\partial_t\langle \vx\rangle=\frac{\I}{\hbar}[H,v_x]=-\frac{2\hbar k}{M} \langle v_\r\rangle.
 \]
Hence, the velocity $\langle v_\r\rangle$ induced by the force $\Fx$ is given by the initial slope of $\langle \vx\rangle$ (\fig{fig_pump}B).
 
The center-of-mass drift $\langle\Delta \r\rangle$, obtained upon integration of $\langle v_\r\rangle$ is shown in \fig{fig_pump}A.  We find that it varies linearly with  the quasi-momentum variation $\Delta q$ (\fig{fig_pump}A), such that the drift per Bloch oscillation cycle reads 
\begin{equation}
 \frac{\langle\Delta \r\rangle}{\Lr}=0.97(5),
\end{equation}
consistent with a unit winding around the cylinder  of circumference $Y$ \cite{Note1}. The rotation along $y$ occuring over a Bloch oscillation cycle is thus quantized, providing a first manifestation of the non-trivial band topology. 

We now characterize the global band topology by implementing  Laughlin's charge pump experiment, and extend the protocol to reveal the local geometrical properties. 
To simulate the  axial magnetic field  used to drive the pump, we interpret  the complex phase $\varphi$ involved in the $y$-hoppings (see equation \ref{eq_Veff}) as the Peierls phase associated with the  field $\Bbold_\parallel$ threading the cylinder with a flux 
\begin{equation}\label{eq_Phi}
\Phi_\parallel=\frac{3\varphi}{2\pi}\Phi_0.
\end{equation}
We vary $\Phi_\parallel$ by adjusting the phase difference $\varphi$ between the laser electric fields involved in the spin transitions.

We drive the pump by slowly ramping the phase $\varphi$, and measure the induced shift of the center-of-mass along the real dimension $x$. The experiment is performed for various values of the quasi-momentum $\langle q\rangle$ uniformly spanning the magnetic Brillouin zone. The $q$-averaged drift, shown in  \fig{fig_pump}C, is consistent with a linear variation
\[
 \frac{\langle\Delta x\rangle}{\lmag}=\mathcal{C}\frac{\Phi_\parallel}{\Phi_0},\quad\mathcal C=1.00(4),
\]
in agreement with the expected quantization of transport by the Chern number $\mathcal C=1$. The pump adiabaticity is checked by repeating the experiment for various speeds of the flux ramp, and measuring identical responses for slow enough ramps \cite{Note1}. 

Our experiments also give access to the anomalous drift of individual momentum states $\Delta x=\Omega(q)\varphi
$, proportional to the  Berry curvature $\Omega(q)$ that quantifies the local geometrical properties of quantum states \cite{lu_geometrical_2016}. As shown in  \fig{fig_pump}D, the measured Berry curvature is flat within error bars, consistent with theory, which predicts  $\Omega(q)=1/(2k)$ with negligible $q$ variation. The  flatness of the Berry curvature is a consequence of the continuous translation symmetry along $x$, making our system similar to continuous two-dimensional systems with flat Landau levels. In contrast, discrete  lattice systems, such as Hofsdtater and Haldane models \cite{hofstadter_energy_1976,haldane_model_1988}, or previous implementations of synthetic Hall cylinders \cite{han_band_2019,li2018bose,liang_coherence_2021}, exhibit dispersive bands with inhomogeneous Berry curvatures.

We have shown that implementing a quantum Hall cylinder gives direct access to the underlying topology of Bloch bands. Our realization of Laughlin's pump protocol could be generalized to interacting atomic systems, which are expected to form strongly correlated topological states of matter  at low temperature. In particular,  at fractional fillings, one expects the occurrence of charge density waves as one-dimensional precursors of two-dimensional fractional quantum Hall states  \cite{tao_fractional_1983}.
The pumped charge would then be quantized to a rational value, revealing the charge fractionnalization of elementary excitations  \cite{laughlin_nobel_1999}.

%

\section*{Acknowledgments}
We thank Jean Dalibard for insightful discussions and careful reading of the manuscript, and Thomas Chalopin for discussions at an early stage of this work.  \textbf{Funding:} This work is supported by  European Union (grant
 TOPODY 756722 from the European Research Council).  \textbf{Author contributions:} 
A.F., J.B.B. and T.S. carried out the experiments, supervised by R.L. and S.N.  All authors were involved in the data analysis and contributed to the manuscript.

\cleardoublepage

\setcounter{equation}{0}
\setcounter{figure}{0}
\setcounter{table}{0}
\setcounter{page}{1}
\makeatletter
\renewcommand{\theequation}{S\arabic{equation}}
\renewcommand{\thefigure}{S\arabic{figure}}

\section*{Materials and methods}

\section{Implementation of spin couplings}

To implement the spin-orbit coupling, we apply a magnetic field along $z$ that induces a Zeeman splitting $\deltaZ=2\pi\times \SI{401(2)}{\kilo\hertz}$ between the successive magnetic levels $m$. The spin dynamics is driven by two-photon optical transitions, using a pair of laser beams as shown in \fig{fig_model}A. Each beam is linearly polarized, along $\ebold_1=\cos\theta\,\ebold_z+\sin\theta\,\ebold_y$ and $\ebold_2=\cos\theta\,\ebold_z-\sin\theta\,\ebold_y$ for the laser beams 1 and 2, which propagate along $\ebold_x$ and $-\ebold_x$, respectively. Their waist $w\simeq\SI{60}{\micro\meter}$ is much larger than the rms size $\sigma\simeq \SI{3}{\micro\meter}$ of the atomic gas, such that the light intensity can be considered uniform.

The laser frequencies are set close to the atomic resonance of wavelength $\lambda=\SI{626.1}{\nano\meter}$ (red detuning from resonance $\Delta\simeq-2\pi\times \SI{22}{GHz}$),  coupling the electronic ground state of angular momentum $J=8$ to an excited level with $J'=J+1$. The proximity to an isolated optical transition leads to spin-dependent light shifts, which we use here to induce resonant spin transitions. Each beam produces a quadratic energy shift of the magnetic sub-levels proportional to $(3\cos^2\theta-1)m^2$ \cite{cohen-tannoudji_experimental_1972}, which we cancel by setting the polarization angle  to $\theta=\acos(1/\sqrt3)\simeq\SI{55}{\degree}$. 

In order to induce the first- and second-order spin transitions, the laser 2 is  monochromatic at frequency $\omega_2$, while the laser 1  has two frequency components $\omega_{1a}$ and $\omega_{1b}$ (\fig{fig_model}A). The component $\omega_{1a}$ is set close to $\omega_2-\deltaZ$, leading to the process ($a$) that drives a spin transition $m\rightarrow m+1$ while imparting a velocity recoil $-2\vrec$, where $\vrec=\hbar k/M$ is the one-photon recoil velocity.  The other frequency component $\omega_{1b}$ is close to $\omega_2+2\deltaZ$, inducing the second process ($b$) with a spin transition $m\rightarrow m-2$ and a recoil $-2\vrec$ (\fig{fig_model}B). Our experiments are performed with spin coupling amplitudes  $t_a=11.5(3)\Er$ and $t_b=7.1(2)\Er$.


\section{Band structure and topology}

\subsection{Band structure}



\begin{figure}[!t!]
\begin{center}
\includegraphics[
 trim={0mm 5mm 0 0.cm},
]{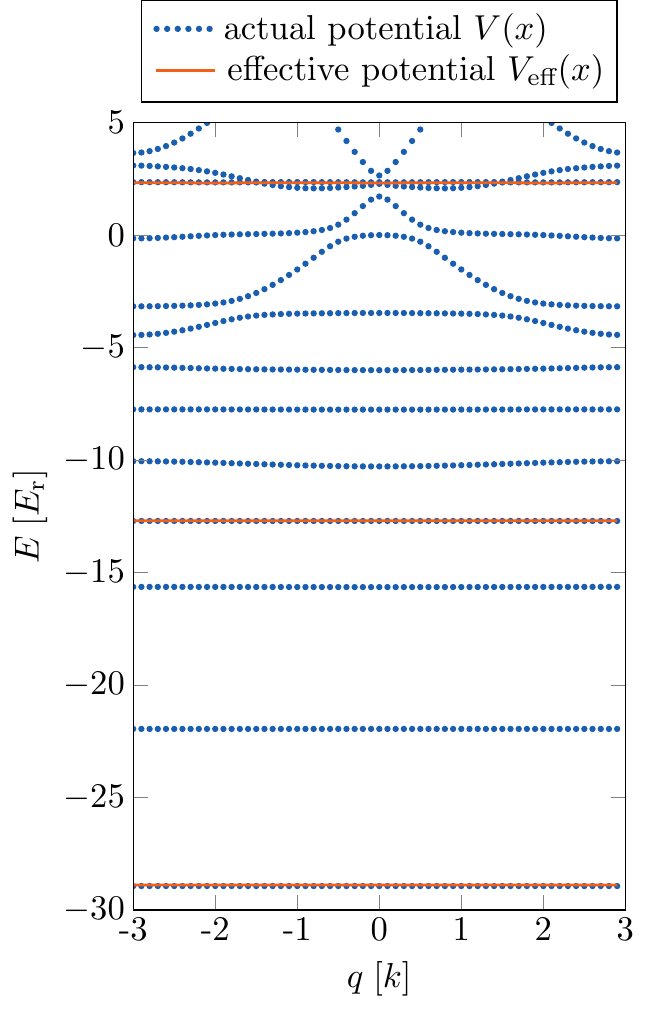}
\end{center}
\caption{\textbf{Band structure.} Band structure calculated for couplings $t_a=11.5\,\Er$ and $t_b=7.1\,\Er$ as chosen in the experiments (blue dots), compared with the band structure of the effective model with a hopping amplitude $t=t_a+t_b=18.6\,\Er$ (red line).
\label{fig_bandstructure}}
\end{figure}

We show in \fig{fig_bandstructure} the structure of magnetic Bloch bands calculated for the couplings $t_a=11.5\,\Er$ and $t_b=7.1\,\Er$ used in our experiment, with either the actual potential $V(x)$ or the effective potential $\Veff(x)$ involving the three spin states $\ket{\r}$ only. The band structure of the effective model matches well several bands of $V(x)$, including the ground band. The bands that are not reproduced correspond to excitations of the system outside the three $\ket{y}$ spin states. They manifest the existence of an additional degree of freedom aside from the $x$ and $y$ dynamics, which could be used in future experiments to engineer more complex topological systems \cite{fabre_2021}. This degree of freedom remains frozen in our experiments, and it plays no role in the interpretation of our results.


\subsection{Topological character of the ground band}

We recall the expression of the Hamiltonian
\begin{align}
H&=\frac{1}{2}M v_x^2+V(x),\label{eq_H_supp}\\
V(x)&=-\E^{\I(\varphi-2 kx)}\left(t_aJ_++t_bJ_-^2\right)+\hc \label{eq_V_supp}
\end{align}
describing the atom dynamics.
The quasi-momentum $q$ being a conserved quantity, the dynamics can be reduced to a Hamiltonian $H(q,\varphi)$ parametrized by the couple $(q,\varphi)$, which varies on the torus $-3k\leq q<3k$ and $0\leq\varphi<2\pi/3$. The  topological character of the ground band is determined by the value of the Chern number \cite{thouless_quantized_1982,luo_tunable_2020}
\[
 \mathcal{C}=\frac{1}{2\pi}\int_{-3k}^{3k}\dd q\int_{0}^{2\pi/3}\dd\varphi\,\Omega(q,\varphi),
\]
where we introduce the Berry curvature
\[
\Omega(q,\varphi)=\I\sum_{n\geq1}\frac{\langle\psi_{0,q,\varphi}|v_x|\psi_{n,q,\varphi}\rangle \langle\psi_{n,q,\varphi}|v_y|\psi_{0,q,\varphi}\rangle-\hc}{(E_{0,q,\varphi}-E_{n,q,\varphi})^2},
\]
with the velocities $v_x=\partial_q H$ and $v_y=\partial_\varphi H$, and we introduce the Bloch state $|\psi_{n,q,\varphi}\rangle$ of the band $n$, of energy $E_{n,q,\varphi}$. The ground band corresponds to the index  $n=0$. In our system, Bloch states of the same band $n$ and quasi-momentum $q$, but different $\varphi$, can be mapped on each other by a spatial translation. Hence, they share the same Berry curvature, which thus only depends on $q$. Integrating over  $\varphi$, we get the relation used in the main text
\[
 \mathcal{C}=\frac{1}{3}\int_{-3k}^{3k}\dd q\,\Omega(q).
\]
For the couplings $t_a$ and $t_b$ used in the experiment, the Berry curvature is extremely flat, equal to $1/(2k)$ for all momenta with a relative variation less than $10^{-4}$.  Its integral over the Brillouin zone yields a Chern number $\mathcal{C}=1$.

\section{Loading of the ground band}

\begin{figure}[!t!]
\begin{center}
 \includegraphics[
 draft=false,scale=1,
 trim={0mm 5mm 0 0.cm},
]{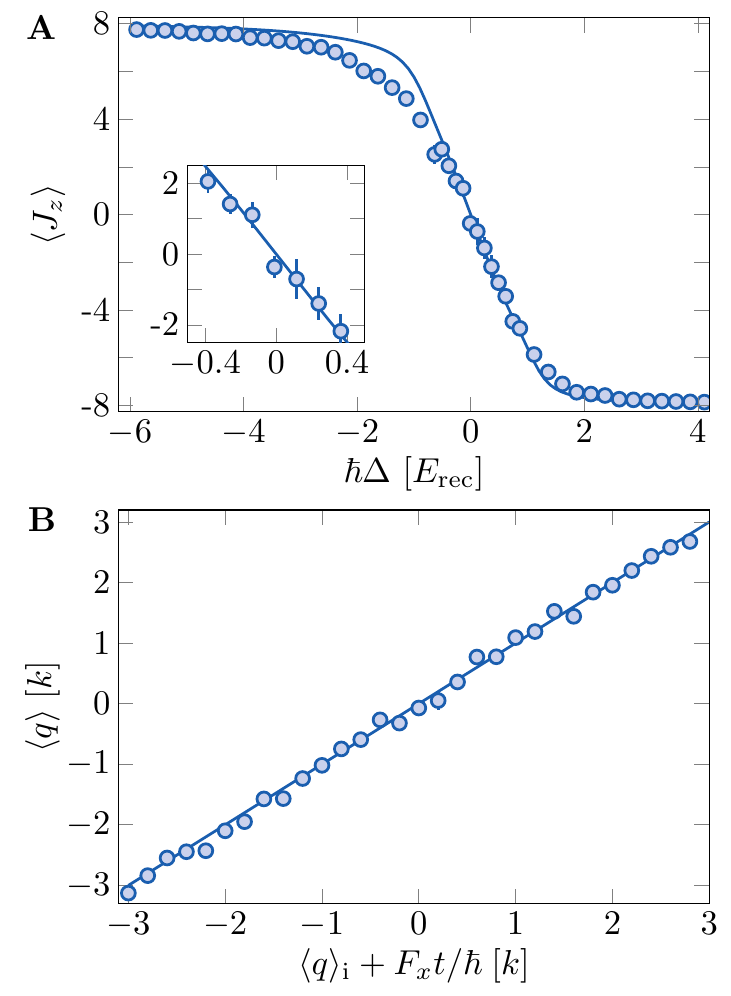}
\end{center}
\caption{\textbf{State preparation.}
(\textbf{A}) Magnetization $\langle J_z\rangle$ as a function of the Zeeman field $\Delta$  for a fixed quasi-momentum $\langle q\rangle\simeq2 \kL$. The solid line is the expected magnetization for the ground state of the Hamiltonian \eqP{eq_H_with_Delta}. The inset highlights the central region around $\Delta=0$. (\textbf{B}) Evolution of the mean quasi-momentum $\langle q \rangle$ during a Bloch oscillation, compared to the expected law $\langle q \rangle = \langle q \rangle_i + F_x t/\hbar \Mod{6\kL} $.
\label{fig_SuppControl}}
\end{figure}

\begin{figure*}[h!t!]
\begin{center}
 \includegraphics[
 draft=false,scale=1,
 trim={0mm 5mm 0 0.cm},
]{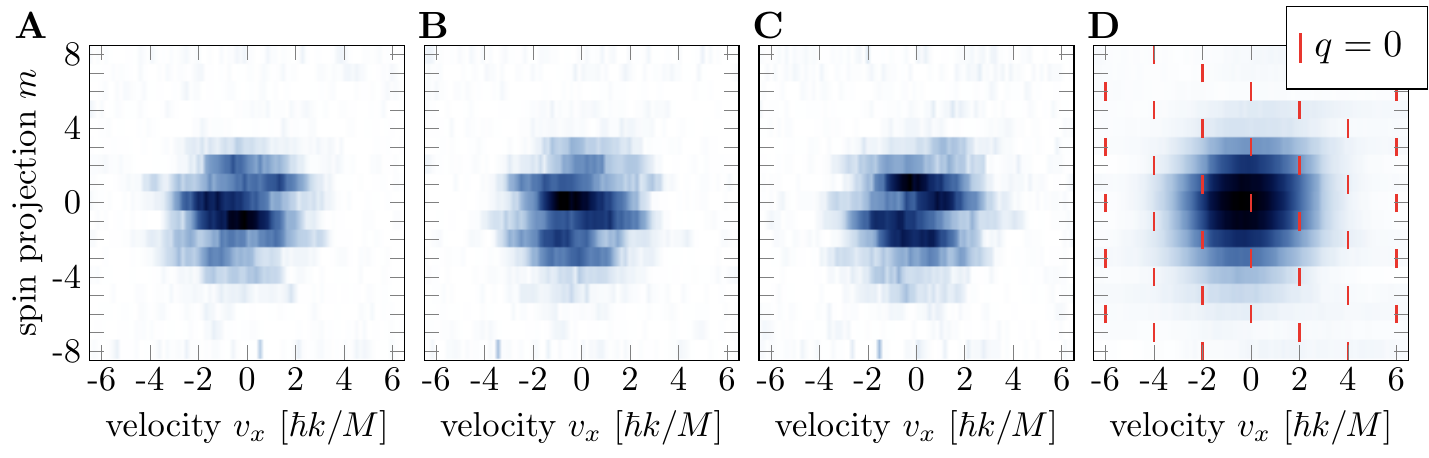}
\end{center}
\caption{\textbf{Averaging and deconvolution of velocity distributions.} 
\textbf{A},\textbf{B},\textbf{C.} Spin-resolved velocity distributions measured for $\langle q\rangle\simeq-2k,0,2k$ (A,B,C, respectively).  \textbf{D.} Velocity distribution averaged over $\langle q\rangle$. The contribution from a given $q$ can be extracted by selecting the velocity components given by \eqP{eq_velocity_q} (example of $q=0$ shown as red lines).
\label{fig_q_deconvolve}}
\end{figure*}

\subsection{Control of the Zeeman field and quasi-momentum}

The preparation of Bloch states of the ground band with a given quasi-momentum $q$ uses additional control parameters provided by setting the laser frequencies away from the Zeeman resonance conditions.  We define the detunings $\delta_{2a} = \omega_{2a} - (\omega_1-\deltaZ)$ and $\delta_{2b} = \omega_{2b} - (\omega_1+2\deltaZ)$. The time dependency of the Hamiltonian can be suppressed with the right choice of transformations. First, we consider the atom motion in   a reference frame moving at velocity $v^\ast = (\delta_{2a} + 2 \delta_{2b})/(6\kL)$ with respect to the laboratory frame. Second, we apply a gauge transform defined by the unitary operator $U = \E^{\I (2 \kL x - (\deltaZ - \Delta)t) J_z}$ with $\Delta = (\delta_{2a} - \delta_{2b})/3$. Using the rotating wave approximation to suppress  fast-oscillating terms, we obtain a static Hamiltonian 
\begin{align}
H&=\frac{1}{2}Mv_x^2+V(x)+\hbar\Delta J_z,\label{eq_H_with_Delta}
\end{align}
where $V(x)$ is defined in \eqP{eq_V} in the main text and $\Delta$ plays the role of a Zeeman field. When  $\Delta$ is set to zero, this Hamiltonian reduces to the one  considered in the main text, given by \eqP{eq_H}.
A time-dependent frame velocity $v^\ast$ results in an inertial force $F_x$ along the real dimension $x$, which we use to drive Bloch oscillations in our system.

\subsection{Protocol for the ground-band loading}

The preparation of the ground state of the Hamiltonian \eqP{eq_H} with $t_a=11.5(3)\Er$ and $t_b=7.1(2)\Er$ is realized as follows. We set the initial laser frequencies such that the frame velocity $v^\ast$ cancels, and the Zeeman field is set to  $\Delta = 16\,\Er/\hbar$. This value is large enough to ensure that the gas is  almost fully polarized in  $\ket{m=-J}$. The atoms have a zero mean velocity $\langle v_x\rangle=0$, such that the mean quasi-momentum reads $\langle q\rangle_{\text i}=M\langle v_x\rangle/\hbar+2\kL\langle J_z\rangle \Mod{6\kL}=2\kL$. 

To load the ground band of the desired Hamiltonian with $\Delta=0$, we first increase the light intensities to their final values in $\SI{100}{\micro\second}$. We then ramp the Zeeman field $\Delta$ towards zero in $\SI{500}{\micro\second}$. This ramp duration is a compromise to ensure adiabaticity while minimizing spin-changing collisions occuring on the timescale of a few milliseconds. The minimum value of the gap to the first excited band $\Delta E_{\text{min}}\simeq5\,\Er$ sets the timescale for adiabaticity $\tau=\hbar/\Delta E_{\text{min}}\simeq\SI{10}{\micro\second}$, much shorter that the chosen ramp duration. We confirm the ramp adiabaticity using a numerical simulation of the atom dynamics, which predicts an overlap with the ground band after the ramp above 97\%.  

The adiabatic loading is checked by probing the system as a function of $\Delta$. For $\Delta\neq0$, we expect the system to exhibit a non-zero magnetization $\langle J_z \rangle$. Its measurement, shown in \fig{fig_SuppControl}A,  agrees well with theory.

The final step in the state preparation is the application of  a force $F_x$ to induce Bloch oscillations. The mean quasi-momentum evolves as  $\langle q\rangle(t) = \langle q\rangle_{\text i} + F_x t/\hbar \Mod{6\kL}$, which we use to prepare the desired quasi-momentum state. We show in \fig{fig_SuppControl}B the measured values of quasi-momentum during a Bloch oscillation, which agrees well with the expected variation. Bloch oscillations are also used to study the topological Hall response, by measuring the velocity along the synthetic dimension $y$ induced by the force $\Fx$ (see main text).

\section{Characterization of the ground band}

The study of the ground band properties are based on the measurement of spin-resolved velocity distributions. For a Bloch state of quasi-momentum $q$, the velocity takes discrete values only, at 
\begin{equation}\label{eq_velocity_q}
v_x=\frac{\hbar}{M}(q-2km+6kp),\quad p\;\text{integer}.
\end{equation}
In our system, the thermal broadening of momentum leads to a continuous velocity distribution (\fig{fig_q_deconvolve}A,B,C). Importantly, the equation \eqP{eq_velocity_q} shows that different quasi-momentum states contribute to distinct velocities in the spin-resolved velocity distribution. The thermal broadening can thus be deconvolved, leading to the velocity and spin distributions resolved in quasi-momentum shown in \fig{fig_ground_band}B,C. In practice, in order to treat all quasi-momenta on equal footing, we first average the spin-resolved velocity distributions measured for various values of $\langle q \rangle$ uniformly spanning the first Brillouin zone. We then deconvolve the data by selecting the velocity components of a given $q$ from the averaged distribution, according to \eqP{eq_velocity_q} (\fig{fig_q_deconvolve}D).

\section{Transverse response in Bloch oscillation experiments}

\begin{figure}[!t!]
\begin{center}
 \includegraphics[
 draft=false,scale=1,
 trim={0mm 5mm 0 0.cm},
]{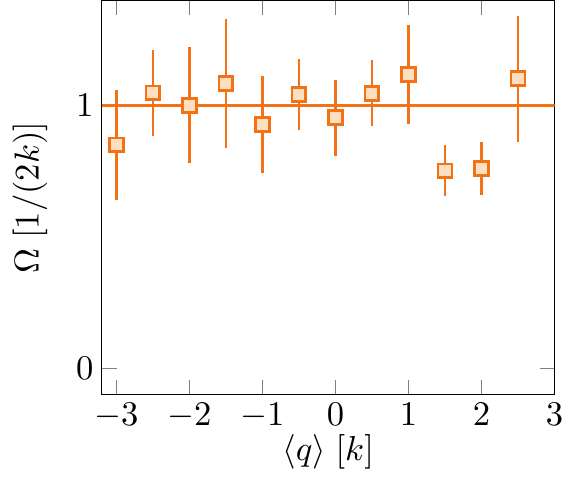}
\end{center}
\caption{\textbf{Berry curvature from Bloch oscillation experiments.} 
Berry curvature measured from the $y$-velocity induced by a force $\Fx$ (orange squares), compared to the expected value $\Omega(q)=1/(2k)$ (solid line).
\label{fig_Omega}}
\end{figure}

We discuss in the main text the adiabatic $y$-drift occuring during a Bloch oscillation driven by a force $\Fx$. In the weak force limit, one expects the mean velocity $\langle v_y\rangle$ to be proportional to the force and to the Berry curvature $\Omega(q)$, as 
\begin{equation}\label{eq_Berry}
\langle v_y\rangle=\Omega(q) F_x/\hbar.
\end{equation}
We show in \fig{fig_Omega} the measured  Berry curvature for various values of the mean quasi-momentum $\langle q\rangle$. The measurements are consistent with a flat Berry curvature -- similar to the measurements with the Laughlin pump protocol, albeit with larger error bars here due to the differentiation operation used to extract the velocity $\langle v_y\rangle$ from the $x$-velocity oscillations (\fig{fig_pump}B). The adiabaticity criterion required for the linear relation \eqP{eq_Berry} to apply is discussed in section \ref{Bloch_oscillation_adiabaticity}.

The adiabatic $y$-drift acquired for a duration $T$ reads
\begin{align}
\langle \Delta y(t)\rangle&=\int_0^T\dd t\,\langle v_y\rangle(t)\\
&=\int_{q_i}^{q(T)} \dd q\,\Omega(q),
\end{align}
where we used $q(t)=q_i+\Fx t/\hbar$. The drift accumulated over a period thus reads
\begin{align}
\langle \Delta y\rangle&=\int_{-3k}^{3k} \dd q\,\Omega(q)\\
&=\Lr\,\mathcal{C},
\end{align}
where $\mathcal{C}$ is the Chern number. This expression  links the quantization of the rotation along $y$ during a Bloch oscillation to the Chern number characterizing the ground-band topology.

\section{Adiabaticity of topological pumps}

The quantization of topological pumps requires the pump control parameters to be varied adiabatically. We present here a study of adiabaticity of the two topological pumps considered in the main text.

\begin{figure}[!t!]
\begin{center}
 \includegraphics[
 draft=false,scale=1,
 trim={0mm 5mm 0 0.cm},
]{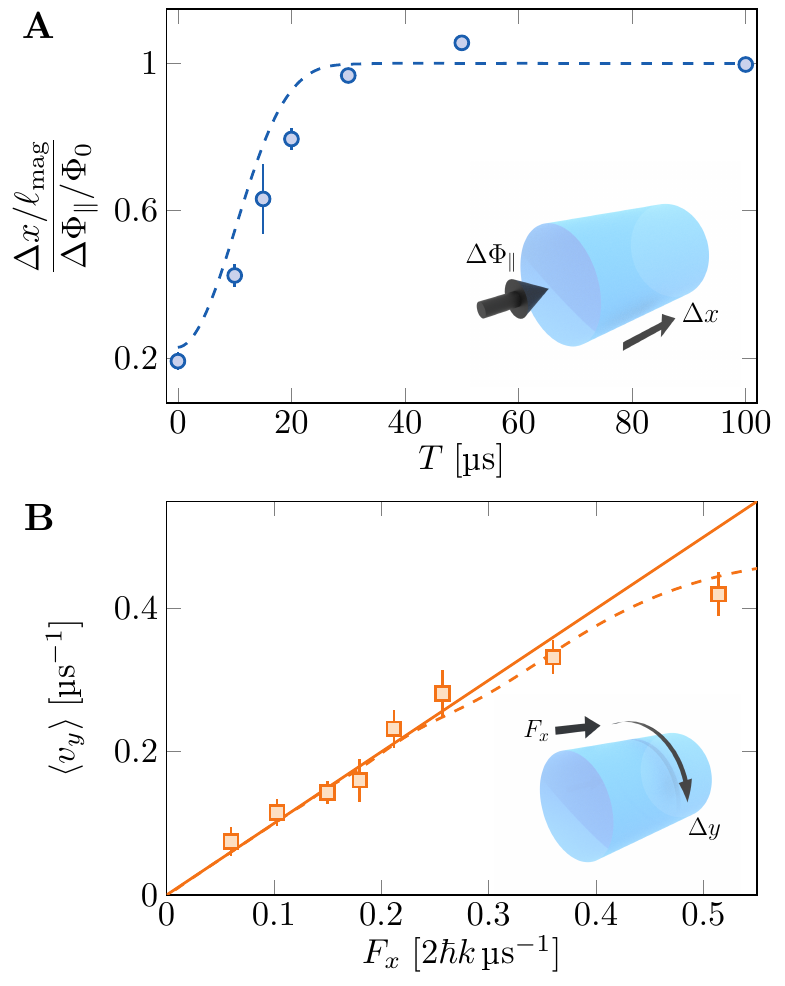}
\end{center}
\caption{\textbf{Adiabaticity of topological pumps.}
(\textbf{A}) Drift of the atomic cloud induced by a flux insertion  $\Delta\Phi_\parallel \simeq 83\, \Phi_0$ as a function of the duration $T$ used to reach the pump rate  $\dot\Phi_\parallel=0.41\,\Phi_0/\si{\micro\second}$. The dashed line is obtained by a numerical simulation of the atom dynamics. 
(\textbf{B}) Velocity $\langle v_y \rangle$ along the synthetic dimension measured as a function of the applied force $F_x$. The solid line is the expected linear relation expected at small forces, and the dashed line is obtained by a numerical simulation. 
\label{fig_Pump_adiabaticity}}
\end{figure}

\subsection{Laughlin pump adiabaticity}
 The Laughlin pump is driven by inserting a longitudinal magnetic flux $\Phi_\parallel$. The flux is increased linearly in time  at a rate $\dot{\Phi}_\parallel$, after a ramp-up phase of the rate using an s-shaped profile of duration $T$. We show in \fig{fig_Pump_adiabaticity}A the mean atom displacement $\langle\Delta x\rangle$ as a function of the ramp time $T$. For slow ramps  $T\geq\SI{50}{\micro\second}$, the displacement is compatible with the value given by the Berry curvature $\Omega(q) \simeq 1/(2 \kL)$. Deviations are observed for faster ramps in agreement with a numerical simulation of the atom dynamics. The measurements shown in \fig{fig_pump}A are performed with a ramp duration $T=\SI{100}{\micro\second}$ in the adiabatic regime.

\subsection{Bloch oscillation adiabaticity\label{Bloch_oscillation_adiabaticity}}

The other topological pump studied in our work consists in the motion along the synthetic dimension $y$ induced by  a force $F_x$ along the real dimension $x$.  We show in \fig{fig_Pump_adiabaticity}B the mean velocity $\langle v_y\rangle$  as a function of $F_x$. For $F_x\leq0.5\,\hbar k/\si{\micro\second}$, the $y$-velocity varies linearly with $\Fx$, in agreement with the expected adiabatic response. The deviations observed for larger forces are well accounted for by a numerical simulation of the atom dynamics. The measurements shown in \fig{fig_pump}B use a force $\Fx=0.18\hbar k/\si{\micro\second}$, well in the adiabatic regime.

\section{Bandgap measurements}

We studied the low-energy excitations of the system, which are of two types: the excitations described by the effective Hall cylinder model, which assume the spin state to remain in the 3-dimensional  $\ket{y}$ manifold, and the excitations leaving this subspace.

\begin{figure}[!t!]
\begin{center}
 \includegraphics[
 draft=false,scale=1,
 trim={0mm 5mm 0 0.cm},
]{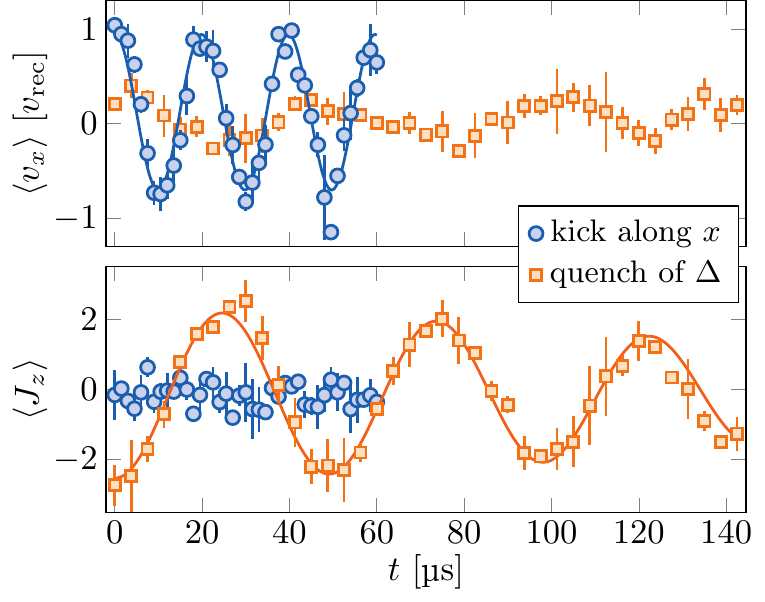}
\end{center}
\caption{
\textbf{Bandgap measurements.}
Variation of mean velocity $\langle v_x\rangle$ and magnetization $\langle J_z\rangle$ as a function of time, following a  kick along $x$ (blue circles) or a quench of the Zeeman field $\Delta$ (orange squares). The blue solid line is a fit of the experimental data with a sine function. The orange solid line is a fit with two sine functions to account for a residual excitation of the second excited band.
\label{fig_Bandgap}}
\end{figure}

In order to probe excitations of the effective Hall cylinder model, we apply a kick along $x$ using a short pulse of the force $\Fx$, which does not affect the spin degree of freedom. As shown in \fig{fig_Bandgap}, we measure an oscillation of the mean velocity $\langle v_x\rangle$, associated to an energy gap of $16.1(1)\Er$, close to the expected value of $15.7\,\Er$ (corresponding to the gap to the third excited band of the full model, see \fig{fig_bandstructure}). During this evolution, the magnetization $\langle J_z\rangle$ remains close to zero, as expected for an excitation within the $\ket{y}$ spin states.

We also studied the excitation to the first excited band, which involves spin states outside the $\ket y$ manifold. To promote the system to this band, we prepare the ground state with a non-zero Zeeman field $\Delta$, such that the system exhibits a non-zero magnetization $\langle J_z\rangle$. We then quench the Zeeman field to zero, and measure the subsequent evolution of the $x$-velocity and magnetization. We measure an oscillation of the magnetization with a longer period, corresponding to a gap of $6.7(1)\,\Er$, close to the expected value of $6.2\,\Er$.

\end{document}